\documentclass[journal]{IEEEtran}

\ifCLASSINFOpdf
\else
\fi
\usepackage{multirow} 
\usepackage{multicol} 
\usepackage{subfig}
\usepackage{algorithmic,algorithm}
\usepackage{graphicx}
\usepackage{amsmath,amssymb,amsfonts}

\begin{document}
%
\title{Cellular Traffic Prediction Using Online Prediction Algorithms}
%
%
%

\author{Hossein~Mehri,~\IEEEmembership{Student~Member,~IEEE,}
        Hao~Chen,~\IEEEmembership{Member,~IEEE,}
        and~Hani~Mehrpouyan,~\IEEEmembership{Member,~IEEE}
}

%
%

\markboth{Journal of \LaTeX\ Class Files,~Vol.~xx, No.~x, January~2024}%
{}
%



\maketitle

\begin{abstract}
The advent of 5G technology promises a paradigm shift in the realm of telecommunications, offering unprecedented speeds and connectivity. However, the efficient management of traffic in 5G networks remains a critical challenge. It is due to the dynamic and heterogeneous nature of network traffic, varying user behaviors, extended network size, and diverse applications, all of which demand highly accurate and adaptable prediction models to optimize network resource allocation and management.
This paper investigates the efficacy of live prediction algorithms for forecasting cellular network traffic in real-time scenarios. We apply two live prediction algorithms on machine learning models, one of which is recently proposed Fast LiveStream Prediction (FLSP) algorithm. We examine the performance of these algorithms under two distinct data gathering methodologies: synchronous, where all network cells report statistics simultaneously, and asynchronous, where reporting occurs across consecutive time slots. Our study delves into the impact of these gathering scenarios on the predictive performance of traffic models. Our study reveals that the FLSP algorithm can halve the required bandwidth for asynchronous data reporting compared to conventional online prediction algorithms, while simultaneously enhancing prediction accuracy and reducing processing load. Additionally, we conduct a thorough analysis of algorithmic complexity and memory requirements across various machine learning models. Through empirical evaluation, we provide insights into the trade-offs inherent in different prediction strategies, offering valuable guidance for network optimization and resource allocation in dynamic environments.
\end{abstract}


%
\IEEEpeerreviewmaketitle

\section{Introduction}
%
%
%
%
\IEEEPARstart{T}{he} proliferation of services offered by 5G and beyond networks, coupled with the widespread use of portable devices, has led to a substantial increase in the demand for wireless networks. Forecasts predict that wireless cellular connectivity will reach 100 billion connections by 2030, as indicated by reports from Huawei~\cite{[577-1]} and Ericsson~\cite{[577-2]}. Simultaneously, global mobile data traffic is anticipated to experience a fourfold increase by the end of 2027~\cite{[577]}.

Mobile operators face numerous challenges in meeting the escalating demands for services and the desire for enhanced Quality of Service (QoS). To tackle these challenges, operators are compelled to implement innovative strategies for the expansion and optimization of cellular networks. A viable solution involves augmenting network density through the deployment of a substantial number of micro-cells around the periphery of macro-cells~\cite{[577]}. This strategy aims to alleviate traffic congestion on macro-base stations by offloading it to micro-base stations, accompanied by load balancing measures. In this context, the softwarization of the control process and the implementation of proactive resource management mechanisms are of paramount importance for optimizing resource allocation. Most of these methods are built upon anticipatory decision-making paradigms~\cite{[579-1]}, with network traffic forecasts considered as crucial information in the context of operation, administration, and management in cellular networks~\cite{[562-1]}. Proactive resource allocation methods offer enhancements not only in the quality of experience (QoE) for users but also the potential to reduce operator costs by improving the energy efficiency of the network. This efficiency can be realized through strategic measures such as toggling the underlying base stations on or off based on predictions of future high or low traffic loads. However, the actual gain of these methods largely depends on the accuracy of the prediction~\cite{[579]}. Moreover, the required processing time and resources are crucial factors in the employment of predictive methods, determining the feasibility of the method's deployment. Proactive management methods are mostly time-critical mechanisms that operate with a live network, requiring continuous and prompt information delivery to function properly. This necessitates the adoption of agile and reliable prediction algorithms in 5G and 6G networks~\cite{[579-1]}.

Providing accurate and continuous predictions in 5G and 6G networks poses a challenging problem due to the complexity of users' traffic patterns, directly influencing the performance of traffic engineering mechanisms. Existing forecasting methods are predominantly tailored for offline scenarios, where prediction models are trained using a set of data, initialized by the latest historical data, generating predictions for a few steps ahead. However, as fresh data arrives from the live network, this process necessitates repetitive updates. These methods typically require a fixed historical window to initialize it before making predictions. Managing and updating this sequential data, particularly as the network size increases, imposes significant processing loads and demands substantial memory. This procedure may be suitable to planning and development applications where rough but long-term predictions are acceptable. On the other hand, proactive resource allocations in live networks demand accurate short-term and continuous predictions for active performance optimization.

Moreover, the traffic on a mobile network can exhibit highly irregular patterns due to factors like device mobility and network heterogeneity. Many prediction algorithms lose accuracy after only a few steps due to the intricate nature of traffic patterns. Conversely, a capable online prediction algorithm can efficiently leverage continuously collected fresh data to enhance prediction accuracy. This motivation prompts the authors of this work to propose an agile and lightweight traffic prediction scheme suitable for live scenarios, eliminating the need to store historical data while effectively utilizing fresh data without imposing excessive processing loads on the system. 

Another challenge in real-world wireless communication systems is the inherent delay and asynchronous collection of the most recent statistics of communication traffic, resulting from delays in sensing, calculation, and reporting. This delay significantly impacts system performance \cite{[562-2]}. To mitigate the effects of delays on performance degradation, the proposed mechanism should forecast future communication using partial information to support system functionalities \cite{[562-3]}.

In this study, we focus on online prediction algorithms that do not require retraining the machine learning network to make accurate predictions. Specifically, we implement two state-of-the-art online prediction algorithms on machine learning models and compare their performance and complexity in predicting the traffic of a live mobile network. Utilizing these algorithms, which operate in an online manner without the need for retraining the model, offers improved tools to achieve remarkable accuracy without the drawbacks associated with traditional online learning algorithms, such as forgetting past information or imposing a heavy processing load for weight updates.

\subsection{Prior Works}
Deep learning-based models are widely employed in traffic prediction problems due to their effectiveness in capturing complex patterns and relationships within large datasets~\cite{[587]}. Compared to statistical models, deep learning-based models exhibit superior performance in prediction accuracy and scalability~\cite{[578]}. In a study by Khalid and Manar~\cite{[581]}, the performance of ARIMA and LSTM models in predicting network traffic in Open Radio Access Network (O-RAN) structure was compared. They state that while LSTM outperformed the ARIMA model in terms of accuracy, the ARIMA model demonstrated better predictions of sudden spikes in the data, highlighting the need for a more comprehensive model, such as a combination of the two, for accurate network traffic prediction.
In another comparison,~\cite{[569]} evaluated the performance of LSTM and GRU models for predicting mobile internet traffic. The authors utilized the K-Means clustering algorithm to group mobile network cells based on their similarities in internet traffic patterns. Then, they adopted an instance of both LSTM and GRU models for each cluster. Their simulations showed that the LSTM model outperformed the GRU in terms of prediction accuracy. However, the study suffered from high data processing complexity.
\cite{[574]} proposed a zero-touch deep-learning-based prediction model that automatically selects the best hyperparameters using genetic algorithms (GA). However, the complexity of the proposed method and the need for careful feature selection and engineering for the GA model were identified as drawbacks.
The authors of~\cite{[585]} utilized transformers and densely connected CNNs for predicting network traffic, achieving more accurate predictions at the expense of increased complexity.

ConvLSTM~\cite{[535]} is considered as a powerful neural network architecture that combines the benefits of convolutional neural networks (CNN) with long short-term memory (LSTM) networks. ConvLSTM's ability to effectively learn both spatial and temporal dependencies, coupled with its relatively low complexity, makes it a popular choice among researchers working on traffic prediction problems~\cite{[580],[562-2],[586]}.
Researchers in~\cite{[562-2]} employed ConvLSTM in parallel with 3D-CNN layers to extract temporal correlations from time-series data. Another study ~\cite{[586]} integrated ConvLSTM into a generative adversarial network (GAN) model, utilizing user statistics to predict network traffic. However, this approach required users to report information such as signal strength, latency, and throughput, making the model vulnerable to issues like noisy data and increased processing and communication load. ~\cite{[580]} combined ConvLSTM with an attention mechanism to improve prediction accuracy by incorporating external information such as the location of base stations, user social behavior, and events affecting traffic patterns. Despite performance improvements, this integration introduced challenges like increased computational complexity, generalization issues, and dependency on the availability of external information on the computing node.
Authors in~\cite{[582]} addressed the data gathering issue in large networks and proposed a federated aggregation mechanism to mitigate this challenge. However, the method has identified drawbacks, including sensitivity to communication delays, scalability issues, and limited robustness to varying network conditions.

Many existing methods primarily focus on enhancing prediction accuracy by employing sophisticated models, resulting in increased computational complexity that may render them unsuitable for practical applications. While integrating external information into predictions is often considered a heuristic solution to enhance accuracy, it can expose models to new challenges such as data inaccuracies in practical situations.
Conversely, operational live networks demand an agile and reliable mechanism that is resilient to data gathering delays and missing information prevalent in practical scenarios. Furthermore, there exists a research gap in considering live scenarios and harnessing the potential of infusing frequently collected fresh data from the network. This research gap motivates our proposal for a mechanism capable of functioning in live networks, leveraging frequently collected data to enhance accuracy without imposing excessive complexity on the system.
\subsection{Contributions}
In this work, we study machine learning-based approaches to predict the traffic of cellular network in live scenarios. Moreover, we study the applicability of Fast LiveStream Prediction (FLSP) algorithm, a recently introduced algorithm in our previous paper~\cite{[myPaper]} for live scenarios, on several types of ML networks such as LSTM, CNN-LSTM, and ConvLSTM and compare them with the modified statistical methods such as ARIMA and SARIMA models. The contributions of this paper can be summarized as follows:
\begin{itemize}
    \item We study the performance of LSTM, CNN-LSTM, and ConvLSTM models in predicting the traffic of a live cellular network and compare them to ARIMA and SARIMA models.
    \item We compared the performance of the FLSP algorithm and rolling algorithm in predicting cellular network traffic in live scenarios.
    \item We modify the existing statistical methods to work in online scenarios for a fair comparison.
    \item In live prediction scenarios, both synchronous and asynchronous data collection is considered in simulations to illustrate the sensitivity of the algorithms to the missing data points.
    \item The complexity of algorithms are extracted for different architectures to provide detailed comparison.
    \item A comprehensive discussion is provided to highlight the advantages of FLSP algorithm over the rolling algorithm in reducing the required bandwidth to collect data in asynchronous data gathering scenarios.
\end{itemize}

\subsection{Organization}
This paper is organized as follows. The traffic prediction problem is formulated as a time-series prediction problem in section~\ref{problemDef}. Online prediction algorithms are explained in section~\ref{onlineAlg}. Section~\ref{predModels} explains the machine learning models explored in this work. Section~\ref{Dataset} is devoted to explanation of real-world dataset and data gathering scenarios. The time complexity and required memory of prediction algorithms for each model is extracted in section~\ref{complexitySec}. Simulation results and discussions are provided in section~\ref{Sim}. Finally, conclusion is provided in section~\ref{Conc}.

\section{Problem Definition}
\label{problemDef}
In this paper, we study the performance of statistical and ML-based methods in time-series prediction problem. For a fair comparison of statistical and ML-based models, we use a real world dataset collected by Telecom Italia~\cite{[data]}. Moreover, we consider live scenarios where fresh data is frequently collected from the network and is used by online prediction algorithms for accurate and continuous predictions. The main challenge of live scenarios is to appropriately assimilate the frequently collected data into the time-series prediction problem and accurately forecast the volume of traffic. Let the traffic data for each period of time be presented as matrix $M_t$ of size $H$ by $W$: 
\begin{equation}
    M_t = \begin{bmatrix}
        m_t^{(1,1)} & \cdots &m_t^{(1,W)}\\ \vdots & \ddots & \vdots \\
        m_t^{(H,1)} & \cdots &m_t^{(H,W)}
    \end{bmatrix},
\end{equation}
where $m_t^{(x,y)}$ measures the traffic volume in a square cell with coordinates $(x,y)$ at time $t$. These points of cellular traffic show both temporal and spatial correlation \cite{[396-8],[396-9]}. Let $T$ be the total number of time periods for which we have traffic data. Then, the entire time-series traffic data can be represented as sequences of matrices $Hist_1 = \{M_1, M_2, ..., M_t, ..., M_T\}$.
Thus, given the historical data from periods $1$ to $T$, the goal is to learn a function $f(\cdot)$ that takes in the historical traffic matrices $Hist$ as input and outputs an estimate of future traffic matrix $\widehat{M}_{T+1}$:
\begin{equation}
    \begin{split}
        &\widehat{M}_{T+1}=f(Hist_1).    
    \end{split}
\end{equation}

The function $f(\cdot)$ can be any appropriate time-series forecasting model among the available models. To develop an effective and accurate prediction function, we need to first process the data and extract the relevant factors that can capture the temporal and spatial patterns of cellular traffic. It is proven that spatial correlation exist among neighbouring cells while distant cells have low or no information about the traffic at the target cell\cite{[585],[561],[1111]}. Therefore, an appropriate model should be able to extract both spatial and temporal correlation of the traffic data in cellular network.

In multi-step prediction scenario, recursive algorithm is employed where the predicted traffic matrices are added to the historical data and then fed to the model to generate next step forecast matrix. In this case the multi-step prediction problem can be written as:
\begin{equation}
    \widehat{M}_{T+i}=f(Hist_i),
\end{equation}
where:
\begin{equation}
    Hist_i=Hist_{i-1} \cup \widehat{M}_{T+i-1}
\end{equation}
is the updated historical data including the previously predicted traffic matrices.

In offline scenarios, where all the information about the time-series is given at first, multi-step prediction problem only relies on the historical data and the estimated data generated in previous steps. Therefore, taken more steps in multi-step prediction problems results in less accurate results which limits the prediction length. On the other hand, in live scenarios, fresh data from the network frequently becomes available, which can be infused to the current historical data and make long-term predictions possible. In the other word, long-term prediction problem in live scenarios is transformed into several short-term prediction problem. In this case, when a new bunch of data is received, the union of this data and the estimated historical data is fed to the prediction function which provides high quality historical data to the prediction model and consequently, yielding into a better results. Equation~\ref{eq5} displays the historical data update equation using the fresh data and prediction equation.
\begin{equation}
    \begin{split}
        &Hist_{k+1}=(Hist_{k} \cup \widehat{M}_{T+k} \cup M'),\\
        &\widehat{M}_{T+k+1}=f(Hist_{k+1}),
        \label{eq5}
    \end{split}
\end{equation}
where $M'=\{M'_{T+1}, ..., M'_{t}, ...,M'_{T+s}\}, s \leq k$ is the recently collected data. Note that at any point of time $t$ in the above union, if there is an actual value collected from the network ($M'_t$), it will replace the estimated value produced by the prediction model ($\widehat{M}_t$). The main challenge of online scenarios is how to appropriately extract and use the new information incorporated within the new data samples according to the predicting algorithm used for forecasting. A suitable solution maximizes the predictions accuracy by fully extracting the information in the new data samples while keeping the processing time and resources minimum. Moreover, practical challenges such as asynchronous data collection may add to the complexities of online scenarios. In this situation, $M'$ matrices may received with delays or with partial data from the network, meaning that some of the information is missing and should be considered in the calculations. The goal of this paper is to investigate the methods of utilizing the fresh data in the machine learning based models while optimizing the processing load as well as required memory.

\section{Online Prediction Algorithms}
\label{onlineAlg}
In this section we briefly describe the two algorithms designed for live scenarios. The first algorithm is rolling algorithms which is extensively used in time-series prediction problems~\cite{[1112],[1113],[1114]}. The other algorithm, FLSP algorithm, is the latest algorithm introduced in our previous work \cite{[myPaper]} which can be applied to the models that use state-based architectures. 
\subsection{Rolling Algorithm}
\label{RollingMethod}
Traditional solutions widely employ Rolling algorithm for live scenarios~\cite{[1113],[1114]}. This algorithm follows a similar strategy to the offline method, where historical data is used to initialize the prediction model and then recursive algorithm is employed to estimate the future traffic matrices. Then, upon arrival of fresh data, it is concatenated with the historical data, and the oldest samples are removed from the historical data. This is done to avoid storing excessive data in memory and keep the buffer size constant. Subsequently, the updated set of data points is fed into the prediction function to generate predictions for future time steps. In this case, equation \ref{eq5} can be rewritten as:
\begin{equation}
    \begin{split}
        &Hist_{k+1}=(\overline{Hist}_{k} \cup \widehat{M}_{T+k} \cup M'),\\
        &\widehat{M}_{T+k+1}=f(Hist_{k+1}),
    \end{split}
\end{equation}
where $\overline{Hist}_k$ is the truncated version of $Hist_k$ where the oldest samples are removed to keep the size of updated historical data ($Hist_{k+1}$) equal to the the buffer size.
Compared to the traditional method where whole the historical data is retained at each step, rolling method needs a fixed-size buffer at each step which requires much less amount of memory to store the historical data. 
Note that at each step only last predicted samples that include new information about the future is retained and is used by the monitoring systems. This procedure is repeated every time that a new data is collected from the network. While the rolling algorithm can generate continuous predictions in live scenarios, it comes with certain inefficiencies. Notably, it necessitates data buffering to manage historical information effectively. Additionally, the concatenation and removal of samples at each step can introduce computational overhead, affecting the prediction speed and overall efficiency of the method. Moreover, resetting the prediction model and initializing it with updated historical data imposes a large burden on the system and makes this method a demanding solution for live scenarios.

\subsection{Fast LiveStream Prediction (FLSP) Algorithm}
\label{FLSP}
\begin{figure*}[tbp]
\centerline{\includegraphics[scale=0.8]{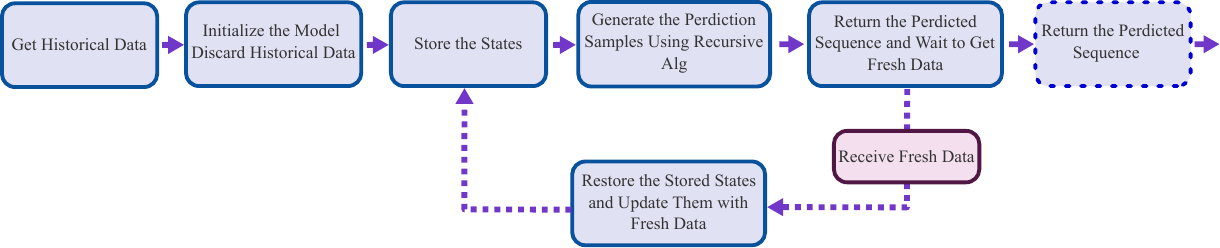}}
\caption{Flowchart of FLSP algorithm.}
\label{fig2}
\end{figure*}

This is the latest algorithm for live scenarios which is proposed in \cite{[myPaper]} and relies on state-based structure of forecasting models. Despite the rolling algorithm, FLSP algorithm stores the latest state of the forecasting model instead of buffering the historical data. In this case, upon arrival of fresh data, FLSP algorithm updates the states of the forecasting model and then generates new predicted samples. The flowchart of this model is depicted in \ref{fig2}. This method offers a lot of advantages over the rolling method. First of all, by storing the model's state we can retain longer dependencies which may lost in rolling method due to the limited size of the buffer. Then, updating the model's state using the fresh data is much faster and less demanding process than updating the historical data and then initializing the forecasting model with it. Finally, storing the states of the model is usually requires less memory than storing the historical data, making the FLSP algorithm a promising replacement of rolling method. The limitation of the FLSP algorithm is that this method is only applicable to the models which utilize a state-based structure, such as RNN, LSTM, GRU, ConvLSTM, and etc. This means we cannot apply this method on statistical models such as ARIMA and SARIMA models. 

\section{Prediction Models}
\label{predModels}
In this paper we investigate two types of forecasting models:
\begin{itemize}
    \item Statistical models (ARIMA, SARIMA);
    \item Deep learning models with memory (LSTM, CNN-LSTM, ConvLSTM).
\end{itemize}
In this section we briefly review the forecasting models and then study their performance in predicting traffic of a cellular network in online model.
\subsection{Statistical Models}
Forecasting time-series data involves using statistical models to predict future values based on past observations. These models are designed to capture and exploit patterns, trends, and seasonality in the time series by fitting a statistical equation to the historical data. There are various statistical models used for time series forecasting, among which, Autoregressive Integrated Moving Average (ARIMA) and Seasonal ARIMA (SARIMA) are two fundamental models.

ARIMA models capture the behavior of time-series data using three components: autoregressive (AR) which accounts for the relationship between current value and past values in the time-series, differencing (I) which is used to make the time-series stationary, and moving average (MA) that models the dependency between the current value and past forecast errors (residuals). Each of these components are explicitly specified in the model as a parameter. A standard notation is used of ARIMA$(p,d,q)$ where the parameters are substituted with integer values to quickly indicate the specific ARIMA model being used. The ARIMA$(p,d,q)$ equation can be abstracted as follows:
\begin{equation}
    \phi(B)\nabla^dy_t=\theta(B)a_t,
\end{equation}
where $B$ is the delay operator, $\nabla$ is difference operator, $\phi(B)$ is auto-regression equation of order $p$, $\theta(B)$ is moving average equation of order $q$, $y_t$ is the past values in time-series, and $a_t$ is past forecast error.

SARIMA is an extension of ARIMA that accounts for seasonality in the data. It includes additional seasonal AR and MA terms to capture repeating patterns. The equation of SARIMA$(p,d,q)(P,D,Q,m)$ can be written as follows:
\begin{equation}
    \phi(B)\nabla^d\varphi_m(B^m)\nabla_m^Dy_t=\theta(B)\vartheta_m(B^m)a_t,
\end{equation}
where $m$ is the seasonality period, $\nabla_m$ is seasonal difference operator, $\varphi_m(B^m)$ is seasonal auto-regression equation of order $P$, and $\vartheta_m(B^m)$ is seasonal moving average equation of order $Q$.

After some simplifications and moving historical terms to the right side of the ARIMA and SARIMA equations, we can have the following forecasting equation:
\begin{equation}
    y_{t}=\Theta(B)a_t-\Phi(B)y_{t-1}=f(A_t,Y_t)
\end{equation}
where $\Phi(B)$ and $\Theta(B)$ are reduced autoregressive and moving average equations, $f(\cdot)$ is the desired forecasting function, $A_t=\{a_0, a_1, \ldots, a_{t}\}$ and $Y_t=\{y_0, y_1, \ldots, y_{t-1}\}$ are the historical error and historical data at time $t$, respectively.

To adjust the ARIMA and SARIMA models for live scenarios, we need to extract the prediction function $f(\cdot)$ from the trained model and then continuously calculate
$A_t$ and $Y_t$ using the fresh data collected from the network. The procedure of online prediction using the ARIMA and SARIMA models is described by algorithm~\ref{alg2}. Note that algorithm~\ref{alg2} falls in the category of rolling methods for live scenarios, where the historical data $A_t$ and $Y_t$ are stored in buffer and upon arrival of fresh data, new samples of historical data are generated using the fresh data and the estimated values.
\renewcommand{\algorithmiccomment}[1]{#1}
\begin{algorithm}
\caption{Online Prediction for RNN-based Networks}
\label{alg1}
\begin{algorithmic}[H]
\STATE Get the $Seed$ data with the length of $l_s$.
\WHILE{$i<=l_s$}
    \STATE calculate $\widehat{y}_i=f(Seed_i)$
\ENDWHILE
\STATE Store the network states.
\WHILE{$i<=S_p$} 
    \STATE calculate $\widehat{y}_{l_s+i}=f(\widehat{y}_{l_s+i-1})$
\ENDWHILE
\STATE Wait to get $Feed$ data with the length of $l_f$. \label{here3}
\STATE Restore the network states.
\WHILE{$i<=l_f$} 
    \STATE calculate $\Bar{y}_i=f(Feed_i)$
\ENDWHILE
\STATE Store network states.
\WHILE{$i<=S_p$} 
    \STATE calculate $\Bar{y}_{l_f+i}=f(\Bar{y}_{l_f+i-1})$
\ENDWHILE
\STATE Add only last $l_f$ steps of $\Bar{y}$ to $\widehat{y}$.
\STATE Return $\widehat{y}$ and go to step~\ref{here3}.
\end{algorithmic}
\end{algorithm}

\renewcommand{\algorithmiccomment}[1]{#1}
\begin{algorithm}
\caption{Online Prediction for Statistical Networks}
\label{alg2}
\begin{algorithmic}[1]
\STATE Get the $Seed$ data with the length of $l_s$.
\STATE Fit the model to the $Seed$ data
\STATE Extract $A_t$, $Y_t$, the reduced AR ($\Phi(B)$) and MA ($\Theta(B)$) equations from the model and create $f(A_t,Y_t)$ function.
\WHILE{$i<=S_p$}
    \STATE calculate: $\widehat{y}_{t}=f(A_t,Y_t)$.
    \STATE add $\widehat{y}_{t}$ and $a_t=0$ to $Y_t$ and $A_t$, respectively.    
\ENDWHILE
\STATE Wait to get $Feed$ data with the length of $l_f$. \label{here2-3}
\STATE Define new AR samples as: $\overline{Y}_t=Y_t \cup Feed$.
\STATE Update $A_t$ with $a_t=Feed(t)-\widehat{y}_t$ for all samples in $Feed$.
\WHILE{$i<=S_p$}
    \STATE calculate: $\overline{y}_{t}=f(A_t,\overline{Y}_t)$
    \STATE add $\overline{y}_{t}$ and $a_t=0$ to $\overline{Y}_t$ and $A_t$, respectively.    
\ENDWHILE
\STATE Add last $l_f$ samples of $\overline{Y}_t$ to $Y_t$.
\STATE Return $Y_t$ and go to step~\ref{here2-3}.
\end{algorithmic}
\end{algorithm}

\subsection{Deep Learning Models with Memory}
\begin{figure*}[htbp]
\subfloat[]{\centerline{\includegraphics[scale=0.8]{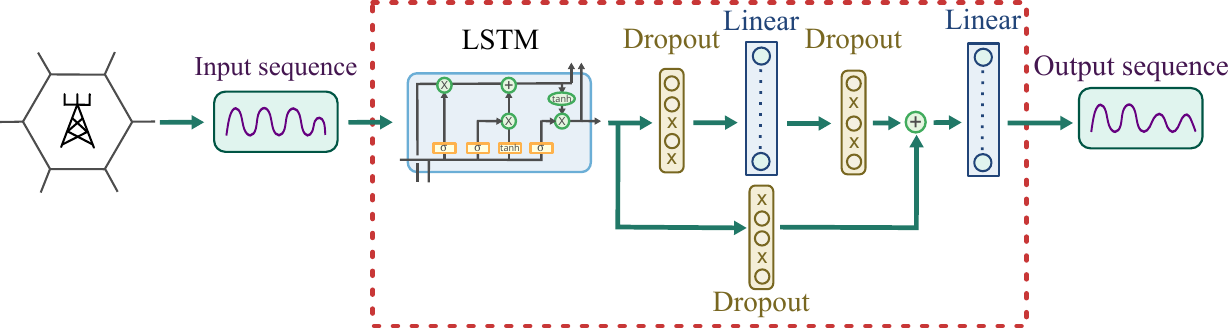}}}

\subfloat[]{\centerline{\includegraphics[scale=0.8]{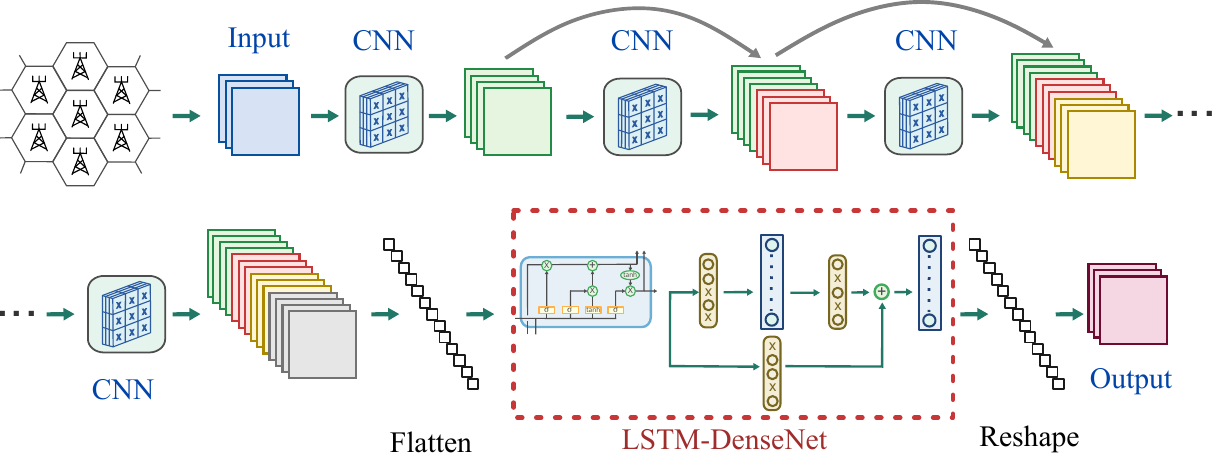}}}

\subfloat[]{\centerline{\includegraphics[scale=0.8]{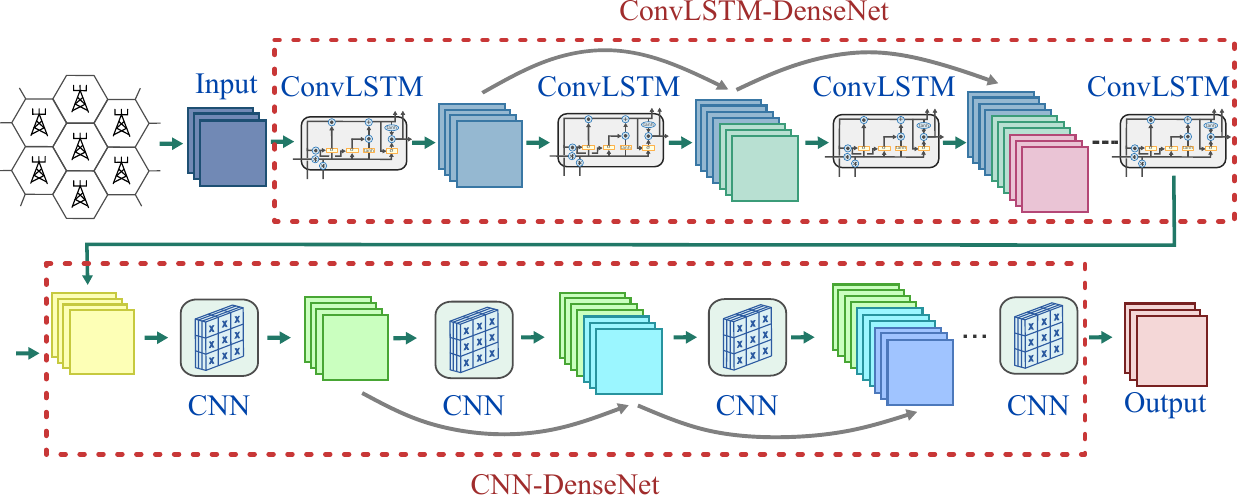}}}
\caption{ML Models with memory: (a) LSTM-DenseNet model, (b) CNN-LSTM model with CNN layers connected in DenseNet architecture, (c) ConvLSTM-CNN model with DenseNet connection between the layers.}
\label{figureX2}
\end{figure*}

Deep learning models with memory play a crucial role in tasks involving time series data, natural language processing, and other sequential data scenarios where understanding context and relationships is essential. Among these, Recurrent Neural Networks (RNN) and their variants, such as Long Short-Term Memory (LSTM), are well-known architectures designed to capture and leverage both long and short-range dependencies. LSTM models, in particular, address the vanishing gradient problem inherent in standard RNNs, which hinders their ability to capture long-term dependencies in sequential data. LSTMs overcome this challenge through a gated architecture that selectively carries relevant information from the past to the future. The states of an LSTM cell efficiently store dependencies between sequential data samples, enabling the implementation of both rolling and FLSP algorithms on LSTM models. In this paper, we employ LSTM models to predict the traffic of individual cells in live mode, applying both rolling and FLSP algorithms for comparison. The LSTM model utilized in this study consists of an LSTM network and a DenseNet network composed of linear layers to predict the traffic of a single cell. The architecture of this network is depicted in Fig.~\ref{figureX2}(a).

Combining Convolutional Neural Network (CNN) and LSTM can be advantageous in analyzing data samples with both temporal and spatial dependencies, such as video frame prediction and cellular network traffic prediction problems. When predicting cellular network traffic, input data matrices can be treated as images and fed into the CNN-LSTM network. The CNN section of the network captures spatial correlations between neighboring cells, while the LSTM section extracts temporal correlations between data samples. Similar to the LSTM architecture, both rolling and FLSP algorithms can be applied to the CNN-LSTM network in live scenarios. Fig.~\ref{figureX2}(b) illustrates the architecture of the CNN-LSTM network.

ConvLSTM, an innovative architecture specifically designed for video frame prediction, takes inspiration from the gated structure of LSTM models, where linear operations are replaced with convolutional ones. This modification enables the network to concurrently learn spatial and temporal patterns. ConvLSTM has shown remarkable predictive capabilities, surpassing existing CNN-LSTM structures, all while maintaining a relatively low level of complexity. Similar to LSTM, ConvLSTM maintains an internal memory cell and various gates to regulate information flow. In this study, the ConvLSTM model is coupled by a CNN section, and layers are connected following the DenseNet architecture, as shown in Fig.~\ref{figureX2}(c).
\captionsetup[subfloat]{labelfont=scriptsize} 

\section{Dataset Description and Data Gathering Scenarios}
\label{Dataset}
In this section we briefly explain the data processing procedure on real-world dataset and the data gathering scenarios in live cellular network.

\subsection{Dataset Description and Data Processing}
\begin{figure}[htbp]
\centerline{\includegraphics[width=0.45\textwidth,height=0.35\textwidth]{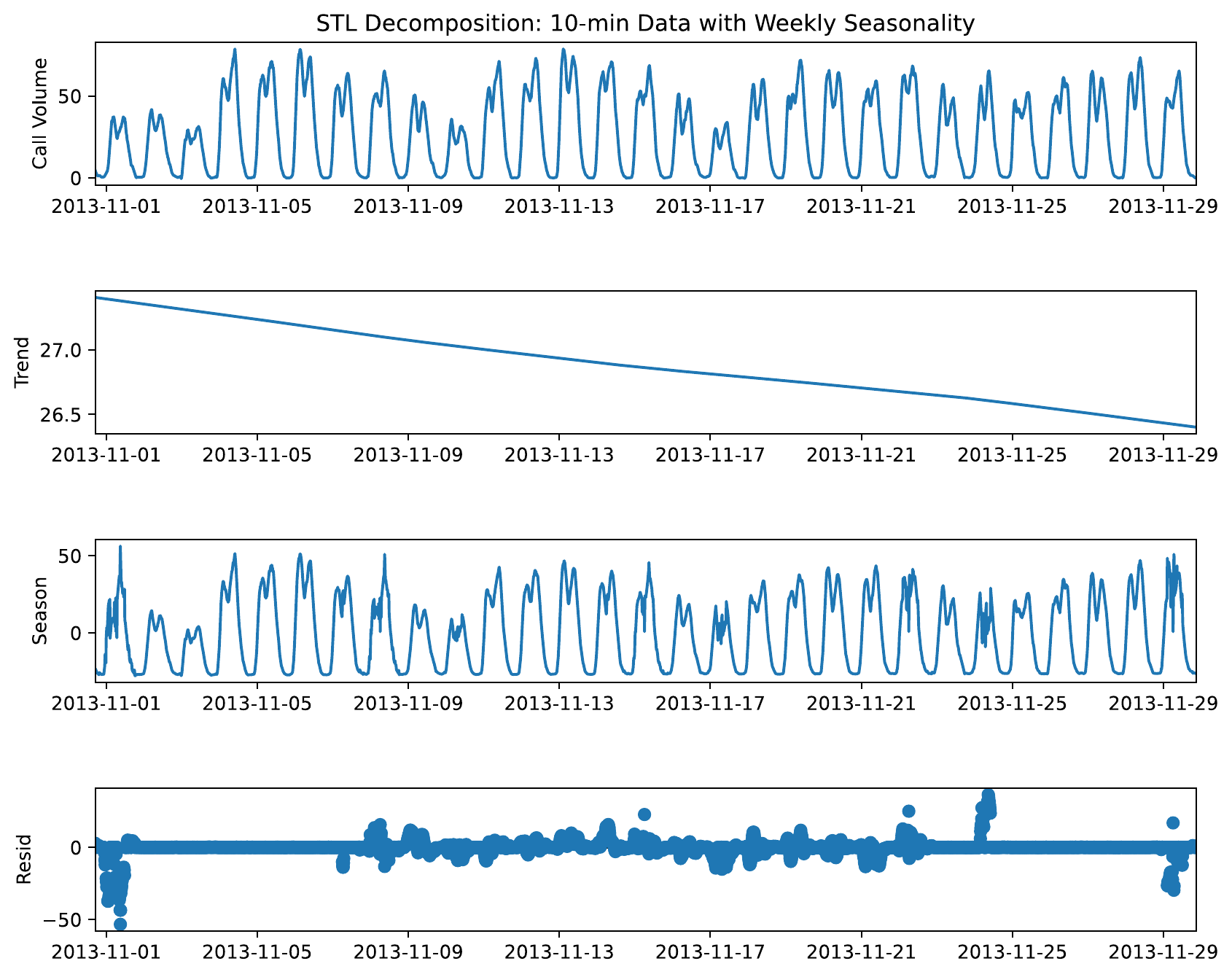}}
\caption{STL decomposition of call volume assuming weekly seasonality.}
\label{figureX5}
\end{figure}

To evaluate the performance of various forecasting models using the online prediction method, we rely on a real-world dataset collected by Telecom Italia in December 2013. This dataset contains Call Detail Records (CDRs) from the cellular network in Milan, Italy~\cite{[data]}. The CDRs provide data on call, SMS, and internet usage volumes within 10-minute intervals, covering users across a grid with 10,000 square cells. In the provided dataset, records are organized by differentiating between incoming and outgoing records, as well as source-destination code pairs. In our analysis, we aggregate the incoming and outgoing traffic from each cell to any destination as a single record. The resulting data takes the form of image-like samples, with dimensions of 100 by 100 by 3. The last dimension represents the volumes of call, SMS, and internet traffic for cells at each time interval. The data exhibits a periodic behaviour in temporal domain. Decomposing the call traffic of a single cell using Seasonal and Trend decomposition using Loess (STL) reveals strong daily and weekly seasonality with a relatively consistent trend. Fig.~\ref{figureX5} illustrates the STL decomposition of an exemplary cell, assuming a weekly seasonality. The residuals from the STL decomposition exhibit noise-like properties, indicating that statistical methods can effectively predict the traffic.

\begin{figure}[htbp]
\centerline{\includegraphics[width=0.4\textwidth,height=0.3\textwidth]{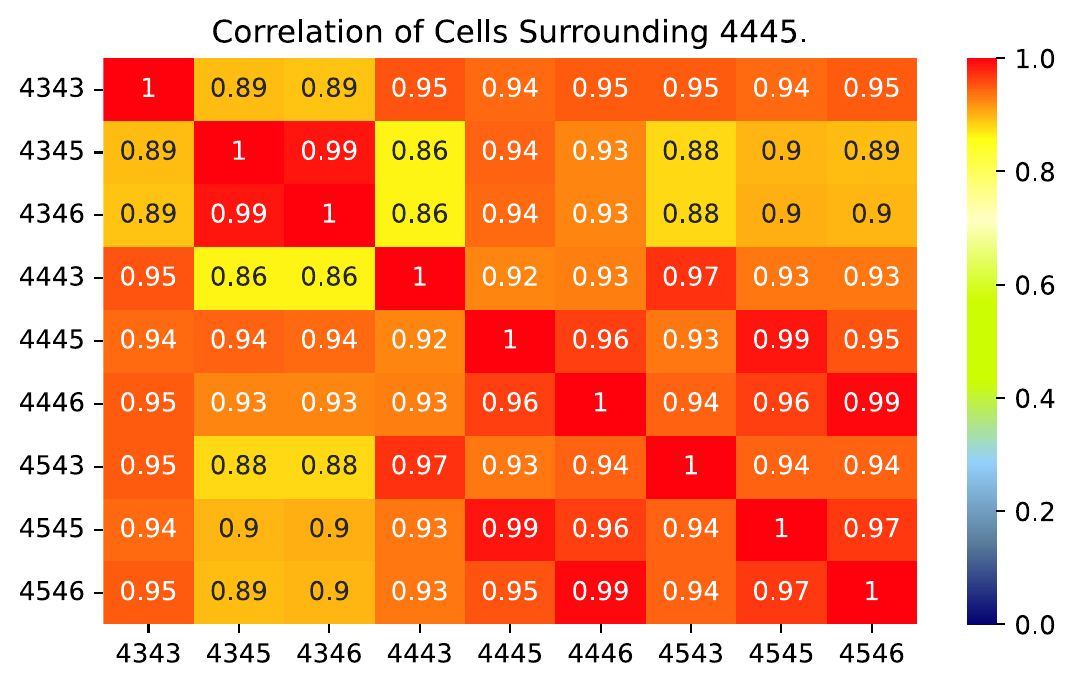}}
\caption{Spatial correlation of cells surrounding cell 4445. It shows high correlation of at least 0.86 among the neighbouring cells.}
\label{figureX6}
\end{figure}

In addition, the data displays a strong spatial correlation among neighboring cells. The correlation heatmap in Fig.~\ref{figureX6} highlights the strong spatial correlation among cells surrounding cell 4445, with a minimum correlation value of 0.86.

In this work, to optimize simulation resources and simplify implementation without compromising generality, we focus on a subset of $900$ cells. This results in input data presented as $(3\times30\times30)$ matrices, where three feature maps correspond to call, SMS, and internet volume, distributed across a 30 by 30 grid.


\subsection{Synchronous and Asynchronous Data Gathering}
\label{Data Gathering}
Data collection in a central node is a crucial technique employed in cellular networks to efficiently gather network traffic data from multiple cells. The simultaneous acquisition of data from all cells within a large scale network presents a significant challenge. On the other hand, asynchronous data collection is a vital technique in cellular networks, allowing for efficient gathering of network traffic data from multiple cells. By capturing data asynchronously, the central node can accommodate variations in timing and volume of data transmission from different cells. This flexibility enables network operators to overcome the challenges of real-time monitoring and synchronization, ensuring that data from all cells can be captured and analyzed effectively. Asynchronous data collection also offers scalability, as the central node can handle data transmission from a large number of cells, facilitating comprehensive network traffic analysis in large and complex cellular deployments. Asynchronous data collection in a central node also ensures that cellular networks can efficiently gather and analyze network traffic data, thereby enhancing the overall quality and efficiency of the network. This makes us to evaluate the performance of the proposed method in asynchronous scenarios.

\section{Computational Complexity and Discussions}
\label{complexitySec}
The computational complexity of a deep learning model can be evaluated from space and time aspects~\cite{[566]}. Here, the number of parameters required to be stored and the number of floating point operations (FLOPs) are chosen as the measures of the apace and time complexity, respectively. Moreover, the extra memory resources required to store the model's state in FLSP algorithm and to buffer historical data in rolling algorithm is calculated for each model. To determine the time complexity of the online prediction approaches, we begin with analysis of time complexity of each model in the forward direction of prediction phase and then study the effects of using FLSP and rolling algorithms on the final complexity. In discussion section, we investigate the superiority of FLSP algorithm in reducing the transmission bandwidth when data is collected in asynchronous scenario.
\subsection{Forward Direction Complexity}
The time complexity of ARIMA and SARIMA models depends on RA and MA components and the number of steps ahead that you want to forecast. For each step ahead, the AR component involves multiplying the past observed values by the corresponding autoregressive coefficients. The MA component involves multiplying the past forecast errors by the corresponding moving average coefficients. For SARIMA model, we should considered the seasonal components as well. Consequently, the time complexity of ARIMA and SARIMA models for one step ahead prediction can be expressed as:
\begin{equation}
\begin{split}
    \mathcal{O}_{arima}&=\mathcal{O}(p+q),\\
    \mathcal{O}_{sarima}&=\mathcal{O}(p+q+P+Q),
\end{split}
\end{equation}
where $p$ and $q$ are the autoregressive and moving average parameters, respectively, and $P$ and $Q$ are the seasonal autoregressive and moving average parameters.
The simplified time complexity of a multi-layer LSTM-DenseNet network is calculated in \cite{[myPaper]} and is in:
\begin{equation}
    \mathcal{O}_{lstm}=\mathcal{O}(L_{lstm}hN+L_Dh^2),
\end{equation}
where $N=max(h,d)$, $h$ denotes the size of the hidden layer of the LSTM, $d$ represents the feature size of the input sequence, and $L_{lstm}$ and $L_D$ show the number of LSTM and DenseNet layers, respectively. 

The time complexity of CNN-LSTM model equals the sum of time complexity of CNN section and LSTM section. The time complexity of a CNN layer with kernel size of $k_l$, input and output feature maps of size $c_{in}$ and $c_{out}$, respectively, and input image of size $(H\times W)$ is in $\mathcal{O}(k_l^2c_{in}c_{out}HW)$. Thus, the time complexity of a multi layer CNN-LSTM model is in:
\begin{equation}
    \mathcal{O}_{cnn-lstm}=\mathcal{O}(\sum_{l=1}^{L_{cnn}} (k_l^2c_{l-1}c_{l}HW)+L_{lstm}hN+L_Dh^2),
\end{equation}
where $L_{cnn}$ denotes the number of CNN layers, $c_l\in\{c_1, c_2, \dots, c_{L{cnn}}\}$ is the size of output feature map of layer $l$ and $c_0$ is the input image feature map size.

ConvLSTM is an extension of the Long Short-Term Memory (LSTM) network that incorporates convolutional operations in its gates to process spatio-temporal data. The key equations of ConvLSTM are shown in \ref{clstmEq} below, where ``$*$" denotes the convolution operator and ``$\circ$", denotes the Hadamard product:
\begin{equation}
    \label{clstmEq}
    \begin{split}
    i_{t} &=\sigma\left(W_{x i}*X_{t}+W_{h i}*H_{t-1}+W_{c i} \odot C_{t-1}+b_{i}\right) \\
    f_{t} &=\sigma\left(W_{x f}*X_{t}+W_{h f}*H_{t-1}+W_{c f} \odot C_{t-1}+b_{f}\right) \\
    C_{t} &=f_{t} \odot C_{t-1}+i_{t} \odot \tanh \left(W_{x c}*X_{t}+W_{h c}*H_{t-1}+b_{c}\right) \\
    o_{t} &=\sigma\left(W_{x o}*X_{t}+W_{h o}*h_{t-1}+W_{c o} \odot C_{t}+b_{o}\right) \\
    H_{t} &=o_{t} \odot \tanh \left(C_{t}\right)
\end{split}
\end{equation}
The convolutional multiplications in these equations can be categorized into two types: (i) input to state transitions which involves input data $X_{t}$; and (ii) state to stet transitions that involves ConvLSTM state $H_{t-1}$. As we preserve the image shape in all layers of the proposed ConvLSTM model (i.e all hidden states and input/output images have size of $(H\times W)$), all of the convolutional multiplications have time complexity of $\mathcal{O}(k_l^2c_{l-1}c_{l}HW)$. The Hadamard products have the time complexity of $\mathcal{O}(c_lHW)$ and the time complexity of summations and activation functions are considered to be in $\mathcal{O}(1)$. Thus, the time complexity of a multi-layer ConvLSTM network in the forward direction with $L_{clstm}$ layers can be expressed as:
\begin{equation}
\label{clstmEq2}
\begin{split}
    \mathcal{O}_{clstm} &=\mathcal{O}(8HW\sum_{l=1}^{L_{clstm}} (k_l^2c_{l-1}c_l)+18HW\sum_{l=1}^{L_{clstm}} (c_l)) \\
    &=\mathcal{O}(HW\sum_{l=1}^{L_{clstm}}(8k_l^2c_{l-1}c_l+18c_l)).
\end{split}
\end{equation}

After removing constants and simplifying equation \ref{clstmEq2}, the time complexity of ConvLSTM can be expressed as:
\begin{equation}
    \mathcal{O}_{clstm} = \mathcal{O}(HW\sum_{l=1}^{L_{clstm}}(k_l^2c_{l-1}c_l)).
\end{equation}
\subsection{Memory Usage and Complexity Analysis for FLSP and Rolling Algorithms}
Assume that the time complexity of employed model is denoted by $\mathcal{O}_{model}$, the time complexity of the model when using FLSP and rolling algorithm can be respectively calculated as \cite{[myPaper]}:
\begin{equation}
    \mathcal{O}_{FLSP}=\mathcal{O}(\frac{l_{f}+l_p}{l_f}\times \mathcal{O}_{model}),
\end{equation}
\begin{equation}
    \mathcal{O}_{Roll}=\mathcal{O}(\frac{l_{buff}+l_f+l_p}{l_f}\times \mathcal{O}_{model}),
\end{equation}
where $l_{buff}$ is the buffer size of rolling algorithm, $l_p$ is the prediction length at each step, and $l_f$ is the size of fresh data collected at each step. Note that the input sequence to the model in rolling algorithm is the union of buffered data of size $l_{buff}$ and fresh data of size $l_f$. Therefore, the proportional complexity of the rolling algorithm to the FLSP algorithm can be expressed as follows:
\begin{equation}
\label{complexity}
    \mathcal{C}=\frac{\mathcal{O}_{Roll}}{\mathcal{O}_{FLSP}}=\mathcal{O}((l_{buff}+l_f+l_p)/(l_{f}+l_p)).
\end{equation}
The required memory to store the states of memory-based deep learning models when using FLSP algorithm is equal to the size of model's hidden states. Thus, the required memory to store hidden states of LSTM and CNN-LSTM is the same and is equal to:
\begin{equation}
    M_{lstm}=M_{cnnlstm}=2 L_{lstm}\ n_h n_{bs},
\end{equation}
where $n_h$ is the number of hidden neurons of an LSTM layer, and $n_{bs}$ is the batch size. Similarly, the required memory to store the states of a ConvLSTM model can be written as:
\begin{equation}
    M_{clstm}=2 n_{bs}\ HW\sum_{l=1}^{L_{clstm}} c_l.
\end{equation}

When using the rolling algorithm, we need to store the historical data of size $b_f$ in the memory. So, for all the models, the required memory can be expressed in terms of buffer size and input data size.
\begin{equation}
    M_{roll}=b_f\ c_0HW,
\end{equation}
where $c_0$ is the input image channels.

ARIMA and SARIMA models require to store the historical autoregressive data ($Y_t$) and historical error data ($A_t$). These sequences of data have a fixed-size and are, respectively, determined by the reduced AR ($\Theta(B)$) and MA ($\Phi(B)$) equations. The size of $\Theta(B)$ is determined by autoregressive parameters of the model and equals to $(p+1)$ and $(P\times m+p+1)$ for ARIMA and SARIMA models, respectively. Similarly, size of $\Theta(B)$ equals to $(q+1)$ and $(Q\times m+q+1)$ for ARIMA and SARIMA models, respectively. Therefore, the required memory by ARIMA and SARIMA models can be expressed as:
\begin{equation}
\begin{split}
    M_{arima}&=(p+q)+1,\\
    M_{sarima}&=(P+Q)m+(p+q)+1.
\end{split}
\end{equation}

Table \ref{table1} summarizes the extracted equations for the complexity and required memory by different models.

\newcommand\T{\rule{0pt}{2.1ex}}       
\newcommand\B{\rule[-0.6ex]{0pt}{0pt}} 
\begin{table*}[!htbp]
\centering
\caption{Prediction complexity of the models and required memory to store the historical data or model's state in rolling and FLSP algorithms.}
\label{table1}
\begin{tabular}{*{4}{|c}|}
\hline
\textbf{Model} & \textbf{Time complexity} & \textbf{Required memory by rolling alg.} & \textbf{Required memory by FLSP alg.}\\
\hline
\textbf{ARIMA} & $\mathcal{O}(p+q)$ & $(p+q)+1$ & - \T \B \\
\hline
\textbf{SARIMA} & $\mathcal{O}(p+q+P+Q)$ & $(P+Q)m+(p+q)+1$ & - \T \B \\
\hline
\textbf{LSTM} & $\mathcal{O}(L_{lstm}hN+L_Dh^2)$ & $b_f\ c_0$ & $2 L_{lstm}\ n_h n_{bs}$ \T \B \\
\hline
\textbf{CNN-LSTM} & $\mathcal{O}(\sum_{l=1}^{L_{cnn}} (k_l^2c_{l-1}c_{l}HW)+L_{lstm}hN+L_Dh^2)$ & $b_f\ c_0HW$ & $2 L_{lstm}\ n_h n_{bs}$ \T \B \\
\hline
\textbf{ConvLSTM} & $\mathcal{O}(HW\sum_{l=1}^{L_{clstm}}(k_l^2c_{l-1}c_l))$ & $b_f\ c_0HW$ & $2 n_{bs}\ HW\sum_{l=1}^{L_{clstm}} c_l$ \T \B \\
\hline
\end{tabular}
\end{table*}

\subsection{Discussion}
In the asynchronous data collection scheme within the mobile cellular network, data gathering by the central node is orchestrated through a systematic allocation of time slots. In this scheme, if the data collection frame is denoted as $f_{collect}$time slots, base stations contribute to the reporting process every other frame. Scheduling in this scenario dictates that half of the base stations submit their data during odd frames, while the remaining stations do so in even frames. Notably, throughout this period, each base station diligently buffers its acquired data, seamlessly transmitting the compiled information within its designated data frame.

In the conventional rolling algorithm, the central node integrates fresh data of size $2\times f_{collect}$ with the buffered historical data. In contrast, the FLSP algorithm focuses solely on the latest $f_{collect}$ data points, disregarding the remainder. Consequently, base stations have the flexibility to discard the oldest half of the data, transmitting only the most recent $f_{collect}$ data points. This streamlined approach not only diminishes the required memory resources in base stations for data storage but also results in report packets that are halved in size. This optimization not only enhances efficiency by reducing memory overhead but also contributes to bandwidth conservation, requiring only half of the bandwidth for transmitting essential data to the central node. This improvement is pivotal in resource-constrained mobile cellular networks, where bandwidth allocation plays a critical role in overall system performance and responsiveness. 

\section{Simulations}
\label{Sim}
The simulation setup is outlined in this section, followed by a discussion of simulation results and subsequent analysis. The Mean Square Error (MSE) function is employed as the performance metric to compare the effectiveness of various approaches.
\subsection{Simulation Setup}
In this paper, we employed two statistical models and three deep learning-based models for predicting cellular network traffic. The best results for statistical based on MSE metric models were achieved using the ARIMA$(3,0,5)$ and SARIMA$(1, 0, 1)(1, 0, 1, 144)$ models, assuming a daily seasonality. Attempting a longer seasonality for the SARIMA model, such as weekly seasonality, proved computationally and memory-intensive, making it impractical for simulation.

The LSTM architecture utilized in simulations is illustrated in Fig.~\ref{figureX2}(a), featuring an LSTM cell with two layers and a hidden size of $1500$, coupled with a DenseNet comprising two linear layers. The dropout value between the layers was set to $0.3$.

The CNN-LSTM model used in simulations, depicted in Fig.~\ref{figureX2}(b), incorporates CNN layers arranged in a DenseNet architecture, followed by an LSTM section with two layers and a hidden size of $1500$. Padding is applied in the CNN section to maintain fixed output feature map sizes.

The last architecture involves two sections, ConvLSTM layers to capture temporal and spatial correlations and CNN layers to generate the output sequence. In this implementation, each section is built up a single block of DenseNet layers, where each layer maintains the feature map size through padding. Fig.~\ref{figureX2}(c) provides a visual representation of the ConvLSTM architecture employed in this study. The first section is consist of $6$ ConvLSTM cells and second section is consists of $5$ CNN layers.

\subsection{Simulation Results}

\begin{figure}
    \centering
    \includegraphics[scale=0.38]{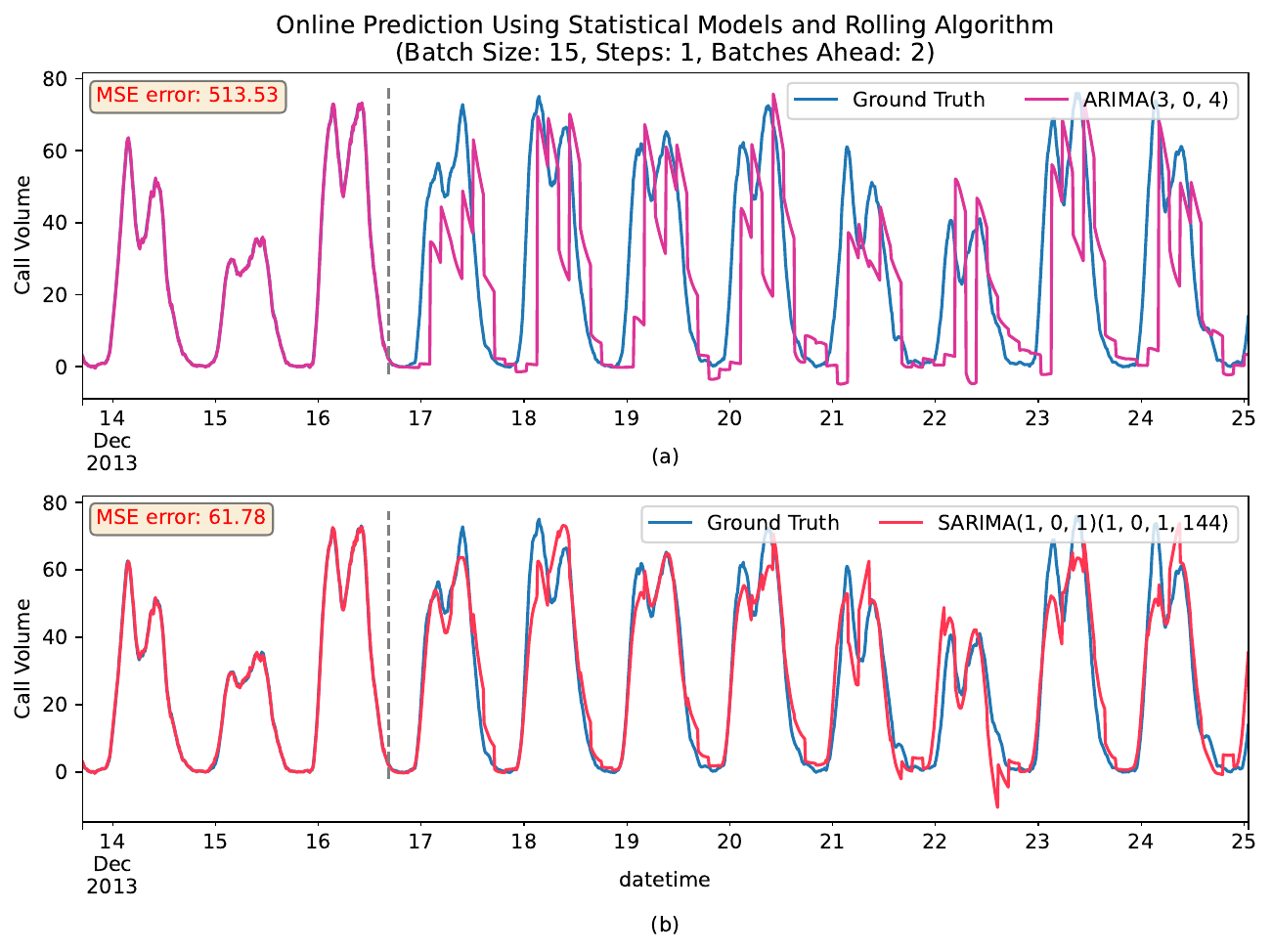}
    \caption{Online predictions using statistical models and rolling algorithm. Fresh data is fed to the models as a single batch of data with a size of $15$ samples, and predictions are made for two batches ahead (with a size of $2*15=30$ time slots) at each step. (a) Comparison of the predicted and ground truth patterns when using ARIMA model, resulting in an MSE of $513.53$. (b) Predicting the traffic pattern using SARIMA model achieves an MSE of $61.78$.}
    \label{figurey1}
\end{figure}

Fig.~\ref{figurey1} illustrates the predicted patterns and ground truth data for statistical models. Simulation results reveal that the SARIMA model, with an MSE value of $61.78$, outperforms the ARIMA model, which has an MSE of $513.53$. This outcome aligns with expectations, as the STL decomposition indicated a seasonal behavior in the data. SARIMA effectively captures this behavior, while the ARIMA model struggles to fully grasp the pattern.

\begin{figure}
    \centering
    \includegraphics[scale=0.37]{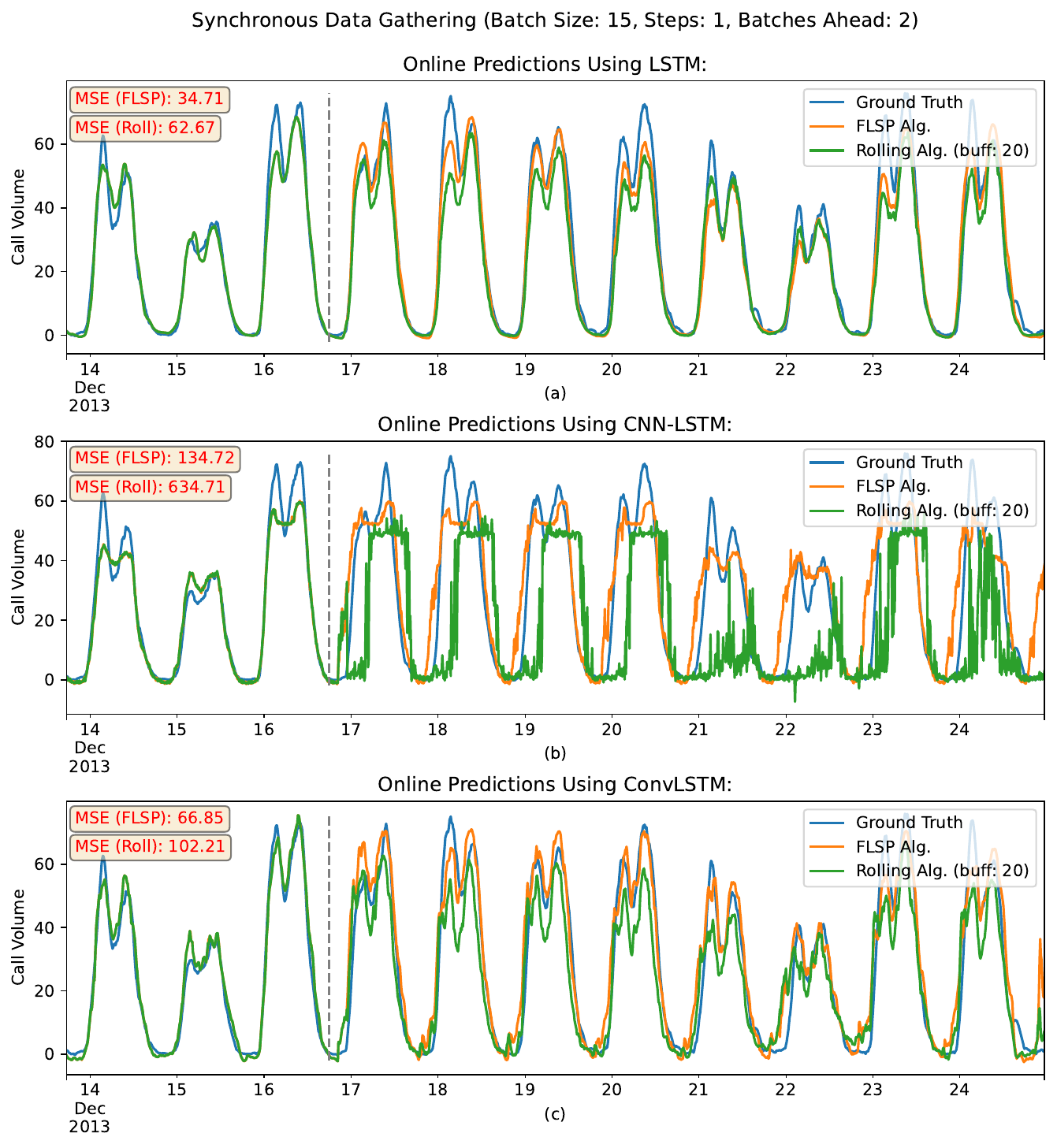}
    \caption{Predicted traffic patterns with synchronous data gathering in live scenarios. In this mode, fresh data is simultaneously collected from all cells every $15$ time slots (one batch size), and predictions are made for next two batches ($2*15=30$ time slots). Predicted traffic patterns using FLSP and rolling algorithm are presented for cell $4445$. (a) Predicting traffic of a single cell using LSTM model: Rolling algorithm with a buffer size of $20$ batches results in MSE of $62.673$, while FLSP algorithm can reach an MSE value of $34.71$. (b) Predicting network traffic using CNN-LSTM model: Predicted pattern of the FLSP algorithm achieves an MSE value of $134.72$, while the rolling algorithm fails to fit the pattern with an MSE value of $634.71$. (c) Predicting network traffic using ConvLSTM model: The ConvLSTM model outperforms CNN-LSTM model with an MSE value of $66.85$ when using FLSP algorithm. This model also shows a better performance when the rolling algorithm is applied with an MSE of $102.21$.}
    \label{figurey2}
\end{figure}

Fig.~\ref{figurey2} illustrates the predicted traffic patterns by deep learning models using two online prediction algorithms. Fig.~\ref{figurey2} (a) depicts the output of LSTM model, showing that the FLSP algorithm improves the prediction accuracy by $50\%$, achieving an MSE of $34.71$. Furthermore, compared to statistical models, the LSTM model exhibits superior performance when using the FLSP algorithm and nearly identical accuracy when using the rolling algorithm with a buffer size of $20$ batches. Similar trends are observed for CNN-LSTM and ConvLSTM models, where the FLSP algorithm significantly outperforms the traditional rolling algorithm.

Asynchronous data gathering is a common procedure in large networks where base stations report their data every other frame. Fig.~\ref{figurey2} depicts the simulation results for an asynchronous data gathering scenario. This can only applied to CNN-LSTM and ConvLSTM models, which attempt to prediction the traffic of all base stations concurrently. In both cases, the FLSP algorithm demonstrate a remarkable performance compared to the rolling algorithm. For a reliable conclusion, simulations are repeated for a hundred times, and the achieved results are summarized in table~\ref{table9x}. In these simulations, buffers of size 5, 10, 20, and 40 batches are adopted for the rolling algorithm. In all cases, the FLSP algorithm outperforms rolling algorithm in terms of accuracy with a remarkable margin while requiring much less simulation time. As we increase the buffer size, the performance of the rolling algorithm approaches that of the FLSP algorithm, but it requires more simulation time. For example, to make prediction for 8 days using ConvLSTM model, the FLSP algorithm requires less than $1$ second, while the rolling algorithm with a buffer size of 20 and 40 requires $6$ and $11$ seconds on an NVidia V100 GPU, respectively. This is compatible with the extracted equations for the complexity of algorithms, where the proportional complexity ratio for buffers of size 20 and 40 using equation~\ref{complexity} returns values of $7.66$ and $14.33$, respectively.

Another advantage of the FLSP algorithm is significant bandwidth saving in asynchronous data gathering scenarios, where base stations only report the latest statistics of size one batch (15 samples), while the rolling algorithm necessitates the base stations to report all the buffered data to the central node, imposing a large transmission load to the network. The FLSP algorithm also outperforms the rolling algorithm in synchronous mode, showing its outstanding ability to predict network traffic while requiring only half of the bandwidth to transfer the collected data. For example, the ConvLSTM model achieves MSE of $130.88$ with the FLSP algorithm in asynchronous mode, which is better than rolling algorithm with buffers of size 5, 10, and 20. 
\begin{figure}
    \centering
    \includegraphics[scale=0.38]{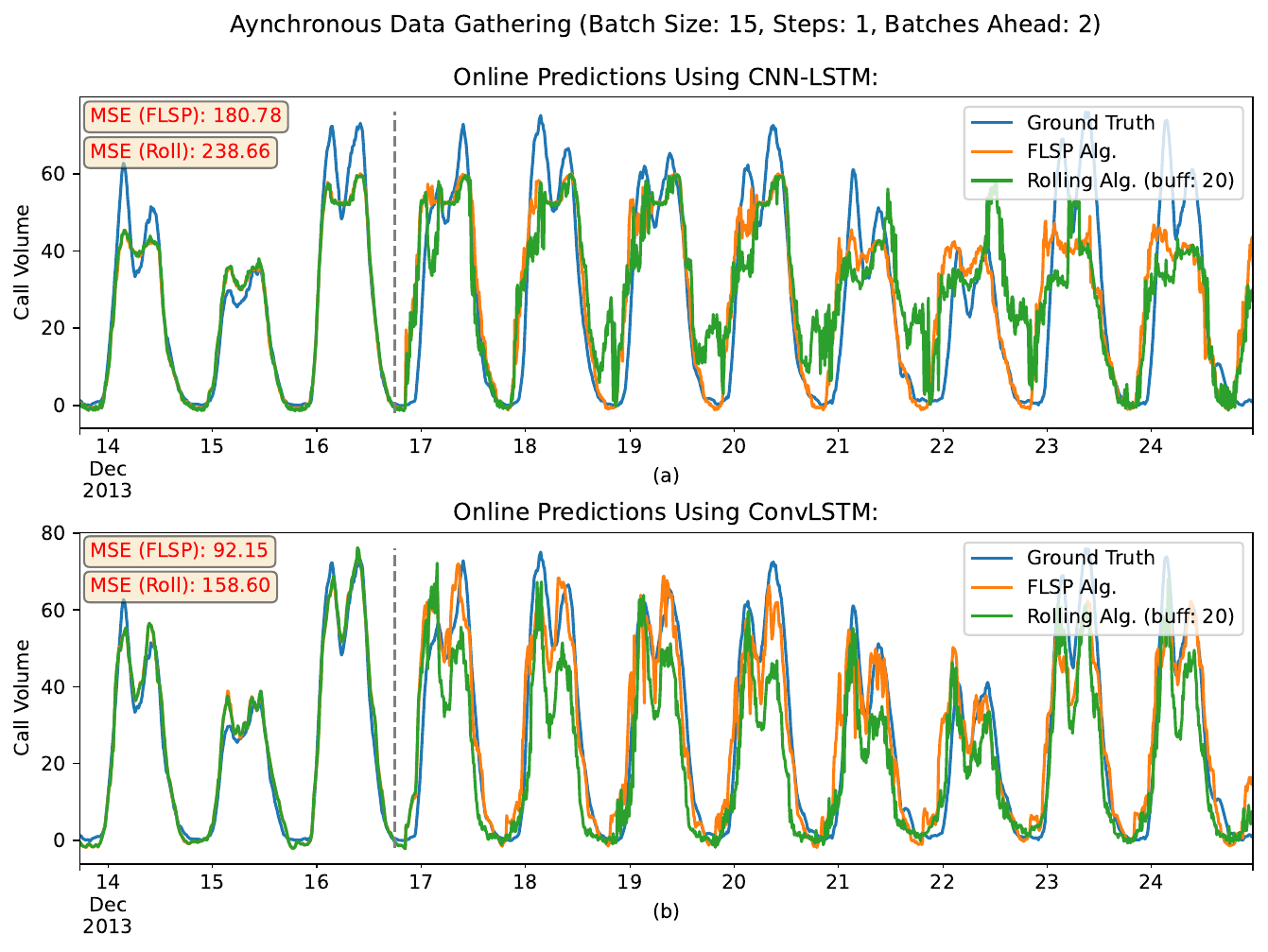}
    \caption{Predicted traffic patterns with asynchronous data gathering in live scenarios. In this mode, base stations report their statistics every other step, i.e. half of the base stations report their data at the current step, and the remaining base stations report their data in the other step. At each step, predictions are made for the next two batches ($2*15=30$ time slots). Predicted traffic patterns using FLSP and the rolling algorithms are presented for cell $4445$. (a) Predicting network traffic using CNN-LSTM model: FLSP algorithm shows better performance than the rolling algorithm. (b) Predicting network traffic using ConvLSTM model: The traffic pattern generated by the FLSP algorithm achieves an MSE of $92.15$, showing better performance than the rolling algorithm with an MSE of $158.60$.}
    \label{figurey5}
\end{figure}
\begin{table*}[!htbp]
\centering
\caption{Performance comparison of the rolling and FLSP algorithm for a hundred repetitions. }
\label{table9x}
\begin{tabular}{*{11}{|c}|}
\hline
\multirow{3}{*}{\textbf{Model}} & \multicolumn{2}{c|}{\textbf{FLSP}} & \multicolumn{8}{c|}{\textbf{Rolling}} \T \B \\
\cline{2-11}
& \multirow{2}{*}{\textbf{Sync}} & \multirow{2}{*}{\textbf{Async}} & \multicolumn{4}{c|}{\textbf{Sync}} & \multicolumn{4}{c|}{\textbf{Async}} \T \B \\
\cline{4-11}
&  &  & \textbf{BS=05} & \textbf{BS=10} & \textbf{BS=20} & \textbf{BS=40} & \textbf{BS=05} & \textbf{BS=10} & \textbf{BS=20} & \textbf{BS=40} \T \B \\
\hline
\textbf{LSTM} & 45.03 & NA & 284.04 & 87.78 & 72.52 & 90.31 & NA & NA & NA & NA  \T \B \\
\hline
\textbf{CNN-LSTM} & 181.08 & 211.51 & 624.34 & 504.50 & 596.03 & 496.68 & 536.15 & 368.09 & 380.38 & 303.09  \T \B \\
\hline
\textbf{ConvLSTM} & 84.17 & 130.88 & 481.78 & 336.79 & 238.06 & 110.16 & 515.83 & 408.11 & 311.14 & 189.44  \T \B \\
\hline
\end{tabular}
\end{table*}

Comparing different deep learning models, the LSTM model achieves better performance compared to the CNN-LSTM and ConvLSTM models. However, it only predicts the traffic of an individual cell, while the other models predict the traffic of all cells simultaneously. The ConvLSTM shows a more accurate and stable performance than the CNN-LSTM model while having fewer parameters. 

Table~\ref{table10} represents the number of parameters and required memory of each model. For brevity and fairness, we only include the buffer size of 40 for the rolling algorithm, which an accuracy close to that of the FLSP algorithm. Among the studied deep learning models, ConvLSTM has the least number of parameters, making this mode the fastest model to train and requires less processing resources to make the predictions. While ConvLSTM requires more memory resources to store its hidden state maps compared to CNN-LSTM model when using FLSP algorithm, it still needs less memory than the rolling algorithm with a buffer size of 40.

\begin{table}[!htbp]
    \centering
    \caption{Complexity and memory analysis.}
    \label{table10}
    \begin{tabular}{|c|c|c|c|}
    \hline
         \multirow{2}{*}{\textbf{Model}} & \multirow{2}{*}{\textbf{Parameters}} &  \multicolumn{2}{c|}{\textbf{Memory}}\\
         \cline{3-4}
         & & \textbf{FLSP} & \textbf{Rolling (buff: 40)} \T \B\\
         \hline
         \textbf{LSTM} & $29,293,502$ & $90,000$ & $1200$  \T \B\\
         \hline
         \textbf{CNN-LSTM} & $36,155,175$ & $90,000$ & $1,080,000$  \T \B\\
         \hline
         \textbf{ConvLSTM} & $235,223$ & $891,000$ & $1,080,000$  \T \B\\
         \hline
    \end{tabular}
\end{table}

\section{Conclusion and Future Work}
\label{Conc}
In conclusion, this paper presents a comprehensive investigation into the prediction of cellular network traffic using various models and algorithms. We considered the less-studied live prediction problem, addressing the challenges of predicting traffic in real-time scenarios. Most existing research in this domain predominantly focuses on offline prediction scenarios, making our exploration into live prediction scenarios a valuable contribution to the field.

We explored statistical models such as ARIMA and SARIMA, as well as deep learning models including LSTM, CNN-LSTM, and ConvLSTM. The study focused on both synchronous and asynchronous data gathering scenarios, evaluating the performance of traditional rolling algorithms against the novel FLSP algorithm. Our results highlight the efficacy of the FLSP algorithm, especially in asynchronous scenarios, where it consistently outperformed the rolling algorithm in terms of accuracy while requiring significantly less simulation time. The bandwidth-saving capabilities of FLSP in asynchronous data gathering further emphasize its practical advantages for large-scale networks.

Deep learning models demonstrated superior predictive capabilities, with ConvLSTM emerging as a promising choice due to its accuracy, stability, and efficient resource utilization. ConvLSTM showed competitive performance with fewer parameters compared to LSTM and CNN-LSTM models. Additionally, the study considered the trade-off between model accuracy and resource requirements, shedding light on ConvLSTM as a fast and resource-efficient model for predicting cellular network traffic.

In summary, this research contributes valuable findings to the field of cellular network traffic prediction, emphasizing the importance of algorithm choice and data gathering strategies, especially in live prediction scenarios—a less-explored aspect where most existing researches predominantly focus on offline prediction scenarios. The presented models and insights can serve as a foundation for developing robust and efficient traffic prediction systems in real-world cellular networks.






\ifCLASSOPTIONcaptionsoff
  \newpage
\fi




\begin{thebibliography}{1}

\bibitem{[577-1]} Huawei Report. Communications Networks 2030. Available: \underline{https://www-file.huawei.com/-/media/corp2020/pdf/giv/industry-reports/}\\\underline{communications\_network\_2030\_en.pdf}, Retrieved June 30, 2022.
\bibitem{[577-2]} Ericsson Mobility Report. Mobile data traffic outlook. \underline{https://www.ericsson.com/en/reports-and-papers/mobility-report/datafore}\\\underline{casts/mobile-traffic-forecast}, Retrieved June 30, 2022.
\bibitem{[577]} E. S. Escriche, S. Vassaki, and G. Peters, ``A comparative study of cellular traffic prediction mechanisms," in \emph{Wireless Networks}, pp.1-19, 2023.
\bibitem{[579-1]} N. Bui, M. Cesana, S. A. Hosseini, Q. Liao, I. Malanchini, and J. Widmer, ``A survey of anticipatory mobile networking: Context-based classification, prediction methodologies, and optimization techniques," in \emph{IEEE Commun. Surveys Tuts.}, vol. 19, no. 3, pp. 1790–1821, 3rd Quart., 2017.
\bibitem{[562-1]} I. Alawe, A. Ksentini, Y. Hadjadj-Aoul, and P. Bertin, ``Improving Traffic Forecasting for 5G Core Network Scalability: A Machine Learning Approach," \emph{IEEE Network}, vol. 32, no. 6, pp. 42–49, 2018.
\bibitem{[579]} G. O. Ferreira, C. Ravazzi, F. Dabbene, G. C. Calafiore, and M. Fiore, ``Forecasting Network Traffic: A Survey and Tutorial with Open-Source Comparative Evaluation," in \emph{IEEE Access}, 2023.
\bibitem{[562-2]} C. Zhang and P. Patras, ``Long-term Mobile Traffic Forecasting Using Deep Spatio-temporal Neural Networks," in \emph{Proc. 18th ACM Int. Symp. Mobile Ad Hoc Netw. Comput.}, pp. 231–240, Jun. 2018.
\bibitem{[562-3]} X. Chen, J. Wang, H. Li, Y. T. Xu, D. Wu, X. Liu, G. Dudek, T. Lee, and I. Park, ``One for All: Traffic Prediction at Heterogeneous 5G Edge with Data-Efficient Transfer Learning," in \emph{2021 IEEE Global Commun. Conf. (GLOBECOM)}, pp. 01–06, 2021.
\bibitem{[587]} X. Wang, Z. Wang, K. Yang, Z. Song, C. Bian, J. Feng, and C. Deng. ``A Survey on Deep Learning for Cellular Traffic Prediction," in \emph{Intell. Comput.}, vol. 3, Jan. 2024.
\bibitem{[578]} W. Jiang, ``Cellular traffic prediction with machine learning: A survey," in \emph{Expert Sys. Appl.}, vol. 201, no. 117163, Sep. 2022.
\bibitem{[581]} K. Ali, and M. Jammal. ``ML-based dynamic scaling and traffic forecasting for 5G O-RAN," In \emph{2023 IEEE Intl. Conf. on Dependable, Autonomic and Secure Computing, Intl. Conf. on Pervasive Intelligence and Computing, Intl. Conf. on Cloud and Big Data Computing, Intl. Conf. on Cyber Science and Technology Congress (DASC/PiCom/CBDCom/CyberSciTech)}, pp. 0444-0451., 2023.
\bibitem{[569]} G. L. Santos, P. Rosati, T. Lynn, J. Kelner, D. Sadok, and P. T. Endo, ``Predicting Short-term Mobile Internet Traffic from Internet Activity using Recurrent Neural Networks," in \emph{Int. J. Netw. Manag.}, vol. 32, no. 3, Oct. 2020.
\bibitem{[574]} J. F. Pajo, G. Kousiouris, D. Kyriazis, R. Bruschi, and F. Davoli, ``Anns going beyond time series forecasting: An urban network perspective," in \emph{IEEE Commun. Mag.}, vol. 59, no. 5, pp. 88–94, 2021.
\bibitem{[585]} B. Gu, J. Zhan, S. Gong, W. Liu, Z. Su, and M. Guizani, ``A Spatial-Temporal Transformer Network for City-Level Cellular Traffic Analysis and Prediction," in \emph{IIEEE Trans. Wireless Commun.}, vol. 22, no. 12, pp. 9412-9423, 2023.
\bibitem{[535]} X. Shi, Z. Chen, H. Wang, D. Yeung, W. Wong, and W. Woo, ``Convolutional LSTM network: A machine learning approach for precipitation nowcasting," in \emph{Adv. Neural Inf. Process. Syst.}, vol. 28, 2015.
\bibitem{[580]} X. Ma, B. Zheng, G. Jiang and L. Liu, ``Cellular Network Traffic Prediction Based on Correlation ConvLSTM and Self-Attention Network," in \emph{IEEE Communications Letters}, vol. 27, no. 7, pp. 1909-1912, July 2023.
\bibitem{[586]} J.-H. Duan, W. Li, X. Zhang and S. Lu, ``Forecasting fine-grained city-scale cellular traffic with sparse crowdsourced measurements," in \emph{Comput. Netw.}, vol. 214, 2022.
\bibitem{[582]} L. Li, Y. Zhao, J. Wang, and C. Zhang, ``Wireless Traffic Prediction Based on a Gradient Similarity Federated Aggregation Algorithm," in \emph{Appl. Sci.}, vol. 13, no. 6, pp. 4036, 2023.
\bibitem{[myPaper]} H.Mehri, H.Mehrpouyan, H. Chen, ``RACH Traffic Prediction in Massive Machine Type Communications," submitted to \emph{IEEE Trans. Machine Learning Comm. Net}.
\bibitem{[data]} Telecom Italia, ``Telecommunications - SMS, Call, Internet - MI," 2015.
\bibitem{[396-8]} T. P. Oliveira, J. S. Barbar, and A. S. Soares, ``Computer network traffic prediction: a comparison between traditional and deep learning neural networks," in \emph{International Journal of Big Data Intelligence}, vol. 3, no. 1, p. 28, 2016.
\bibitem{[396-9]} G. Barlacchi, M. De Nadai, R. Larcher, A. Casella, C. Chitic, G. Torrisi, F. Antonelli, A. Vespignani, A. Pentland, and B. Lepri, ``A multi-source dataset of urban life in the city of Milan and the Province of Trentino," in \emph{Scientific data}, vol. 2, p. 150055, 2015.
\bibitem{[561]} C. Zhang, H. Zhang, D. Yuan and M. Zhang, ``Citywide Cellular Traffic Prediction Based on Densely Connected Convolutional Neural Networks," in \emph{IEEE Commun. Letters}, vol. 22, no. 8, pp. 1656-1659, Aug. 2018.
\bibitem{[1111]} J. Su, H. Cai, Z. Sheng, A.X. Liu, A. Baz, ``Traffic Prediction for 5G: A Deep Learning Approach Based on Lightweight Hybrid Attention Networks," in \emph{Digit. Signal Process.}, Vol. 146, 2024.
\bibitem{[1112]} E. Zivot , J. Wang, ``Rolling Analysis of Time Series," in \emph{Modeling Financial Time Series with S-PLUS}. New York, NY: Springer, 2006, pp.313-360. 
\bibitem{[1113]} Z. Hang, Y. Shi, M. Wen and C. Zhang, ``TBSW: Time-Based Sliding Window Algorithm for Network Traffic Measurement," in \emph{IEEE 17th Int. Conf. Smart City; IEEE 5th Int. Conf. Data Sci. and Syst. (HPCC/SmartCity/DSS)}, pp. 1305-1310, 2019.
\bibitem{[1114]} W. Zhou, D. Wang, H. Li and W. Song, ``Long-Term Forecasting of Time Series Based on Sliding Window Information Granules and Fuzzy Inference System," in \emph{2018 5th Int. Conf. Inf., Cybern., Comput. Social Syst. (ICCSS)}, pp. 375-380, 2018.
\bibitem{[566]}T. Zhou, H. Zhang, B. Ai, C. Xue and L. Liu, ``Deep-learning-based spatial–temporal channel prediction for smart high-speed railway communication networks," in \emph{IEEE Trans. Wireless Commun.}, vol. 21, no. 7, pp. 5333-5345, Jul. 2022.

\end{thebibliography}
%

\begin{IEEEbiography}[{\includegraphics[width=1in,height=1.25in,clip,keepaspectratio]{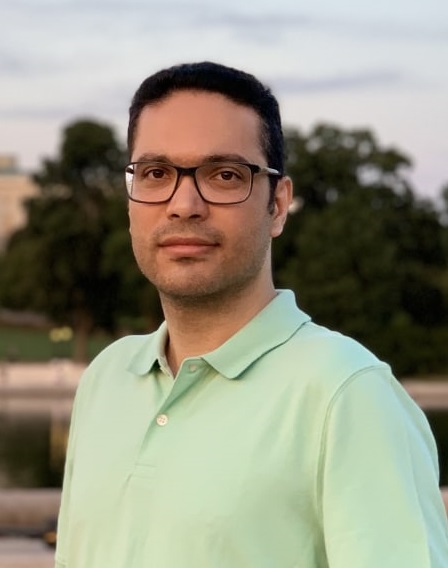}}]
{Hossein Mehri}~(Student member, IEEE)~received the B.Sc. degree from Zanjan National University, Zanjan, Iran, in 2010, and the M.Sc. degree in electrical engineering from Iran University of Science and Technology, Tehran, Iran, in 2013. He is currently pursuing the Ph.D. degree in electrical and computer engineering with Boise State University, Boise, ID, USA. His current research interests include machine learning, wireless communications, and MAC layer protocols.
\end{IEEEbiography}

\begin{IEEEbiography}
[{\includegraphics[width=1in,height=1.25in,clip,keepaspectratio]{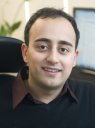}}]
{Hani Mehrpouyan}~(Member, IEEE)~received the B.Sc. degree (Hons.) from Simon Fraser University, Burnaby, Canada, in 2004, and the Ph.D. degree in electrical engineering from Queens University, Kingston, Canada, in 2010. From September 2011 to March 2012, he held a post-doctoral position at the Chalmers University of Technology, where he led the MIMO aspects of the microwave backhauling for next-generation wireless networks project. He was an Assistant Professor at California State University, Bakersfield, CA, USA, from 2012 to 2015, and Boise State University, Boise, ID, USA, from 2015 to 2022. He has received the IEEE Conference on Communication (ICC) Best Paper Award from the Communication Theory Symposium.
\end{IEEEbiography}

\begin{IEEEbiography}
[{\includegraphics[width=1in,height=1.25in,clip,keepaspectratio]{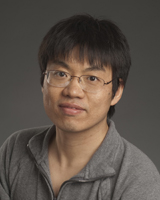}}]
{Hao Chen}~(Member, IEEE)~received the Ph.D. degree in electrical engineering from Syracuse University, Syracuse, NY, USA, in 2007. From 2007 to 2008 and from 2008 to 2010, he was a Postdoctoral Research Associate and Research Assistant Professor with Syracuse University. Since August 2010, he has been an Assistant Professor with the Department of Electrical and Computer Engineering and Computer Science, Boise State University, Boise, ID, USA. His current research interests include statistical signal and image processing, and communications.
\end{IEEEbiography}



\end{document}